\documentclass[twocolumn,tighten]{aastex62}

\hypersetup{colorlinks, citecolor=blue, linkcolor=blue}
\definecolor{fgreen}{rgb}{1,0.6,0.1} 

\newcommand{\trev}[1]{{\textbf {#1}}}

\newcommand{\mne}[1]{$-$}

\received{\today}
\revised{\today}
\accepted{\today}
\submitjournal{ApJ Letter}

\shorttitle{Star-bursting nuclei in old dwarf galaxies}
\shortauthors{Paudel \& Yoon}

\begin{document}

\title{Starbursting Nuclei in Old Dwarf Galaxies}

\author[0000-0003-2922-6866]{Sanjaya Paudel}
\author[0000-0002-1842-4325]{Suk-Jin Yoon}
\affil{Department of Astronomy, Yonsei University, Seoul 03722, Republic of Korea}
\affil{Center for Galaxy Evolution Research, Yonsei University, Seoul 03722, Republic of Korea}
\correspondingauthor{Suk-Jin Yoon}
\email{sjyoon0691@yonsei.ac.kr }

\begin{abstract}
Nuclei of early-type dwarf galaxies (dEs) are usually younger than the galaxy main body, and such discrepancy in age has been a puzzle. To explore the origin of young nuclei in dEs, we study a sample of dEs having compact star-forming blobs that are visually similar to dEs' nuclei but by far bluer. We find that (1) the compact star-forming blobs have a typical stellar mass of one percent of the host galaxy stellar mass; (2) some of the blobs are positioned slightly off from the center of the galaxies; (3) the H$\alpha$ equivalent width measured from the publicly available Sloan Digital Sky Survey fiber spectroscopy shows their formation ages being an order of few Mega-year; and (4) their emission line metallicities, 12\,+\,log(O/H), are as high as the solar value, while the underlying galaxies have the typical stellar populations of dEs, i.e., log(Z/Z$_{\sun}$)\,$\sim$\,$-0.8$. Based on the results, we argue that the central star-forming blobs can provide a caught-in-the-act view of nuclei formation in dEs, and discussing possible formation mechanisms of young nuclei in old dEs. We particularly propose that these off-centered compact star-forming regions may act as seeds of nuclei as proposed in the `wet migration' scenario of \cite{Guillard16}.

\end{abstract}


\section{Introduction}

From the ground- and space-based imaging of early-type dwarf galaxies (dEs), it has been shown that the central nuclei are common in these galaxies \citep{Cote06,Lisker07,Brok14}. In particular, around a stellar mass of 10$^{9}$ M$_{\sun}$, the nucleation fraction reaches $\gtrsim$ 90\,\% \citep{Turner12,Brok14,Janssen19}. These central nuclei have typical sizes of a few to a few tens of parsecs with a mass of nearly 1\,\% of their hosts \citep{Cote06,Rossa06}.

High-resolution imaging from the Hubble Space Telescope observations have revealed that dE nuclei are slightly bluer than the rest part of their host galaxies \citep{Lotz04,Cote06}. Moreover, detailed stellar population studies from integrated light spectroscopy have shown that they are, on average, younger than the host galaxies' stellar halos \citep{Chilingarian09,Paudel11,Guerou15,Kacharov18}. The nuclei of dEs have size and mass ranges similar to those of bright globular clusters and ultra-compact dwarf galaxies, and thus it is natural to speculate that these objects are the remnant nuclei of tidally disrupted dEs \citep{Bekki03,Paudel10,Mieske13,Seth14,Vogge16}.

Nuclei of dEs follow the same scaling relation of a galaxy's mass and its central black hole mass, extending down to the low-mass systems where nuclei are common \citep{Ferrarese06}. In a number of dwarf galaxies, both black holes and nuclei are present at the galactic center \citep{Jiang11,Reines13}. Whether black holes and nuclei form via the same mechanism or either of them serve as a seed for the other is still an open question.

The correlation between the nuclear mass and the host galaxy mass suggests a link between the formation histories of the nuclei and their galaxies, that is, their coevolution. Since the nuclei of galaxies sit at the depth of the galactic potential well, various phenomena such as supermassive black holes, active galactic nuclei, central starbursts, and extreme stellar densities may be associated with the nuclei \citep{Ferrarese06,Boker10,Seth06}. One topic of much recent interest is whether all dEs started with central nuclei and coevolved together, or nuclei formed at the substantially later part of host galaxy evolution \citep{Cen01,Bekki07,Guillard16}. In terms of size, luminosity, and overall frequency, dE nuclei are quite similar to the nuclear star clusters of late-type spirals, allowing us to speculate that dEs are environmentally transformed objects \citep{Cote06,Georgiev14}.

In this context, a detailed understanding of how nuclear clusters formed may reveal the general mechanism that drives the formation of massive central objects of all types. Two basic scenarios have been suggested: (1) nuclear clusters form via mergers of multiple globular clusters accreted through dynamical friction \citep{Tremaine75,Oh00,McLaughlin06,Boker10}, and  (2) nuclear clusters form in situ from gas channeled into the centers of galaxies \citep{Milosavljevic04,Bekki07}. However, in practice, these two formation scenarios do not need to be mutually exclusive. More recently, \cite{Guillard16} proposed a `wet migration' scenario, where a massive star cluster forms near to the center, and it migrates to the center via a combination of interactions with other substructures and dynamical friction. In this work, we present a study of a sample of dEs that possess star-forming blobs near the galaxies' isophotal center. We expect that the study of the central star-forming objects can provide a caught-in-the-act view of nuclei formation in low-mass galaxies.

\section{The Sample}

\begin{figure*}
\centering
\includegraphics[width=18cm]{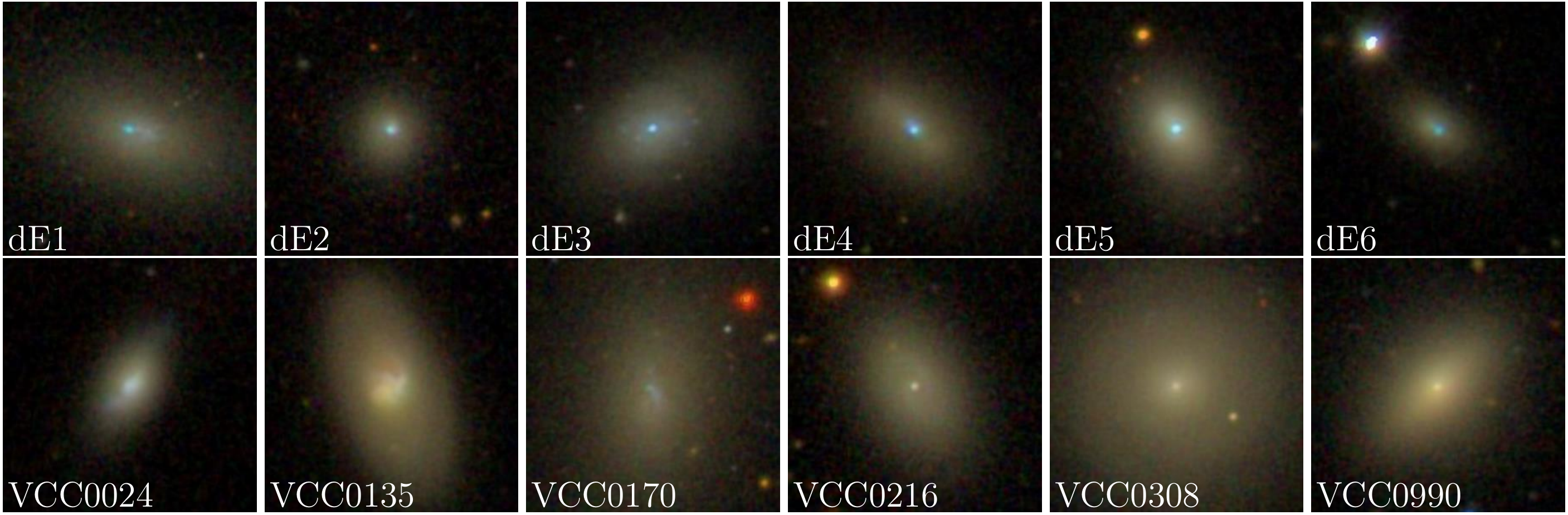}
\caption{$g-r-i$ combined tri-color images obtained from the SDSS sky server with a field of view of 1\arcmin$\times$1\arcmin. The upper row shows our six sample dEs. The lower row shows, in the left, blue centered dEs (VCC0024, VCC0135, and VCC0170) from \cite{Lisker06}, and, in the right, typical dEs with old nuclei (VCC0216, VCC0308, and VCC0990) from \cite{Paudel11}.}
\label{samplefig}
\end{figure*}

The six galaxies analyzed in this work are drawn from the sample of our extended catalog of dEs in the nearby universe (z $<$ 0.01) prepared from the Sloan Digital Sky Survey (SDSS) database \citep[][]{Paudel14}. We select dEs by visual inspection of both the SDSS color images and spectra \citep[see][]{Paudel14}, and these six galaxies in particular draw our attention. Unlike other dEs that contain relatively old or at least non-star-forming nuclei \citep{Paudel11}, they possess a compact star-forming region near the center with no other nuclei, as we can see in the top panels of Figure \ref{samplefig}. These compact star-forming regions mimic young nuclei similar to the star-forming nuclei found in gas-rich spiral galaxies \cite[e.g., NGC 4395;][]{Georgiev14}.

We know that some dEs indeed possess central star formation \cite[e.g.,][]{Lisker06,Urich}. They are commonly known as the blue-centered early-type dwarf (dE\_bc), which possess extended blue regions (see Figure \ref{samplefig}, bottom panels). A standard definition of the nucleus of a dwarf galaxy is the existence of a luminosity excess over the main stellar distribution in the core region \citep{Janssen19}. In this work, we consider the presence of a compact star-forming blob in the core region as a young blue nucleus. This definition may not agree with the conventional definition for relatively old nuclei found in dEs. We are particularly interested in dEs with a compact star-forming blobs because they may be progenitors of young nuclei.

We provide the global properties of our sample galaxies in Table \ref{sampletab}. We have chosen a redshift range of less than 0.005 or distances less than $\sim$20 Mpc. We use this distances limit because we intend to better resolve the compact star-forming regions. In the top panels of Figure \ref{samplefig}, we can see the prominent blue nuclei at the center, contrasting with the red-looking halos of the galaxies. For comparison, we also show some blue-centered dEs of \cite{Lisker06} in the bottom panels. They show more extended star-forming regions at the center, which are not as blue as the cores of our sample galaxies. Our sample galaxies are morphologically typical of dEs, and their $g-r$ colors ranging 0.8\,$\sim$\,1.0 ensure that they are at the red sequence of dwarf galaxies in the color$-$magnitude diagram \citep{Lisker08}.

All galaxies but one are located at the outskirts of the Virgo cluster and have low relative line-of-sight radial velocities with respect to the cluster center galaxy, M87, which makes them likely members of the cluster. We thus consider a mean distance of the Virgo cluster (16.5 Mpc; \cite{Mei07}) as the distance to our galaxies. One, dE6, is located near to NGC 3945 group at a sky-projected distance of 780 kpc from the group center, and the relative line-of-sight radial velocity is as low as 155 km\,s$^{-1}$. We consider a distance of NGC 3945 (21.6 Mpc; \cite{Theureau07}) as the distance to dE6. The data show that none of these are located in a highly dense environment.

\section{Data analysis}

\subsection{Imaging}

The structural and photometric properties of host galaxies and their central nuclei are derived using the archival images available at the SDSS-III Data Archive Server \citep{Abazajian09}. The basic reduction procedures, such as bias-subtraction, flat-field-correction, sky-subtraction, and flux-calibration, were already performed in the archival images. To create an image mosaic with a desired field of view, we use a tool provided by the SDSS-III\footnote{https://dr12.sdss.org/mosaics/}. We subtract the sky-background counts, following the same scheme as in our previous publication \citep{Paudel14}.

For image analysis, we make extensive use of the $i$-band images, with which we derive the galaxies' structural parameters, and use them as the references for other band images. To extract the galaxies' major-axis light profiles, we use the IRAF $ellipse$ task. We first mask all non-related background and foreground objects manually, before running the $ellipse$ task. In the cases of the sizably off-centered nuclei, we also mask them. During the ellipse fit, the center, position angle, and ellipticity are allowed to vary.

\begin{figure*}
\centering
\includegraphics[width=18cm]{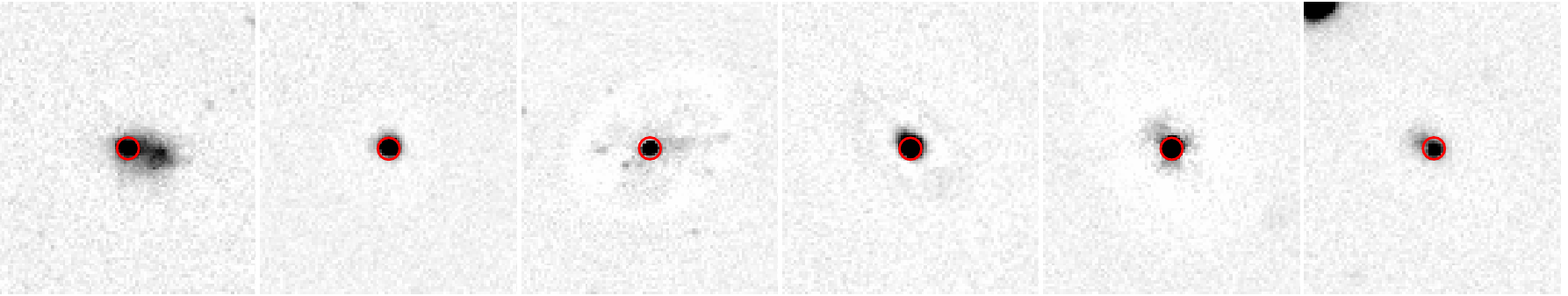}
\includegraphics[width=18cm]{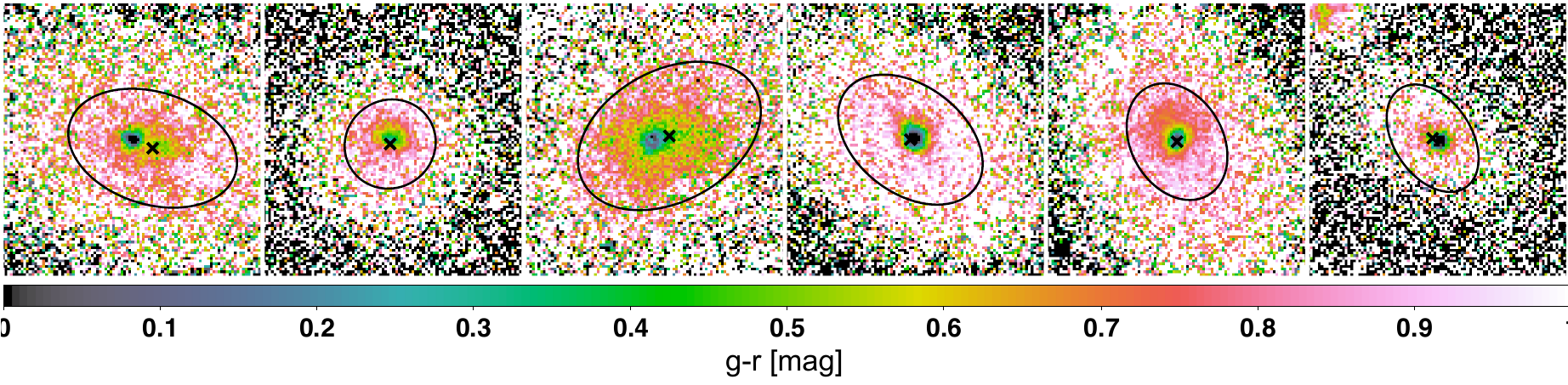}
\caption{({\it Upper row}) The SDSS $g$-band residual map after subtracting best-fit galaxy models (see the text for details). We show only the inner regions with a field a view of 20\arcsec$\times$20\arcsec. The red circles represent the SDSS 3\arcsec{} diameter fibre positions. ({\it Bottom row}) The $g-r$ color map where the galaxies' centers are shown by black crosses and the ellipses represent the best-fit ellipse at half light radii. The field of view is the same as Figure \ref{samplefig}.}
\label{resfig}
\end{figure*}

The observed galaxy light profile is fitted to a S\'ersic function using a $\chi^{2}$-minimization scheme. We skip the central 5\arcsec{} in the galaxy light profile to avoid contamination from nuclear light. The results of the best-fit S\'ersic parameters (i.e., S\'ersic indices and effective radii) are listed in Table \ref{sampletab}, columns 7 and 8, respectively. We show the best-fit model subtracted residual images in the upper panels of Figure \ref{resfig}. The centers of the galaxies are obtained by averaging the centers of the ellipses from $R_{e}$/2 to $R_{e}$. In this range, the ellipse fit shows the center is already stable. To derive the centers of the nuclei, we use the $g$-band residual images with which we calculate their centroids. In the bottom panels of Figure \ref{resfig}, we show the $g-r$ color maps of our galaxies, where the black ellipses represent the best-fit ellipses at the half-light radius. We also identify the centers of the galaxies with black cross marks. The distance between the galaxy center and nuclei are listed in Table \ref{nuctab}, column 5.

We finally measure the total light in the nuclei from the residual images after subtracting the model galaxy light in the SDSS $g-$, $r-$, and $i-$band images. We derive the stellar masses of the nuclei from the $r$-band magnitudes and the mass-to-light ratios obtained from $g-r$ based on \cite{Bell03}.

\subsection{Spectroscopy}

We retrieve the optical spectra of the six dEs from the SDSS data archive. The SDSS spectra were observed with the spectrograph fed with the 3\arcsec{} fibers that correspond to a physical diameter of 240 pc at the distance of the Virgo cluster and 315 pc at the distance of NGC 3945. While carefully matching the positions of nuclei and the fibers, we found that the SDSS fibers were always positioned at the nuclei, rather than the galaxies' geometric centroids (Figure \ref{resfig}, top panels).

The optical spectra of these galaxies show strong emission-lines and resemble a typical spectrum of a star-forming HII region. However, we attempt to derive the properties of the underlying old stellar populations using available information in the absorption features as well as the properties of star formation based on the emission-lines. We first fit the observed galaxy spectrum in a wavelength range from 4000 to 7000\,\AA{} with simple stellar population models from \cite{Vazdekis10}. In doing so, prominent emission-line regions are masked. For this purpose, we use a publicly available full-spectrum fitting tool, the University of Lyon Spectroscopic Analysis Software  \citep[ULySS][]{Koleva09}. The fit is not satisfactory in the cases of dE1 and dE6, which have the strongest H$\alpha$ emission. We caution that the inferred parameters of the underlying old stellar populations should be considered qualitative. To derive them more precisely, the extraction of galaxy spectrum should be done significantly away from the central young nuclei \citep{Paudel11}, where there is no contamination from the emission-lines.

Finally, we measure the emission-line fluxes by analyzing the stellar-continuum-subtracted spectra. In the cases of dE1 and dE6, we directly measure the emission-line fluxes from their original SDSS spectra because, as mentioned above, the fitting of the old stellar population is not satisfactory. We derive oxygen abundances, 12 + log(O/H), using the O3N2 method \citep{Marino13}. The H$\alpha$ emission-line flux is converted to a star-formation rate (SFR) using the calibration of \cite{Kennicutt98}. We find that the Balmer decrement ($c$ = H$\alpha$/H$\beta$) ranges from 3.0 to 3.5, which indicates a mild effect of dust reddening. The estimated values of 12 + log(O/H), SFR and $c$ are listed in Table \ref{nuctab}, columns 7, 8, and 9, respectively.

\section{Results}

The observed colors of nuclei are significantly blue and they all have a negative value of $g-r$. Given that the nuclei have a narrow range of the $M_r$ magnitude ($-12.7$\,$\sim$\,$-13.5$), it is not surprising that their stellar masses of nuclei are similar ($\sim$5$\times$10$^{6}$ M$_{\sun}$). Note, however, that the estimated values of stellar mass are rather uncertain since we use a single optical color ($g-r$) to calculate the mass-to-light ratio \citep{Zhang17}.

In Figure \ref{resfig}, the position of the galaxy center (black cross mark) does not always overlap with the bluest position of the map, suggesting some of the nuclei are off-centered. A half of our sample have the galactocentric distances of the nuclei that are larger than the observational point-spread functions (see Table \ref{nuctab}, columns 5 and 6).

From H$\alpha$, we can get a rough estimate of the star-formation age. For this, we use the Starburst99 model \citep{Leitherer99}, assuming an instantaneous burst of star formation and a solar value of metallicity.  For equivalent width EW(H$\alpha$) higher than 100 \AA{}, the ages of star formation are estimated to be less than 10 Myr. Simple estimation of the stellar mass accumulated over the periods of star formation, assuming a solar value of metallicity, shows that the total stellar mass is two orders of magnitude less compared to the photometrically calculated stellar mass. There are many reasons for this discrepancy. As mentioned above, the use of a single optical color to calculate the mass-to-light ratio may overestimate the stellar mass. A detailed analysis of spectral energy distributions with longer baseline wavelength is required. Besides, the subtraction of underlying host galaxy light might not be perfect. In particular, due to the lack of sufficient spatial resolution, it is not trivial to extrapolate the central light from the model that is mainly fit with the outer part of the galaxies.

\begin{figure*}
\centering
\includegraphics[width=8cm]{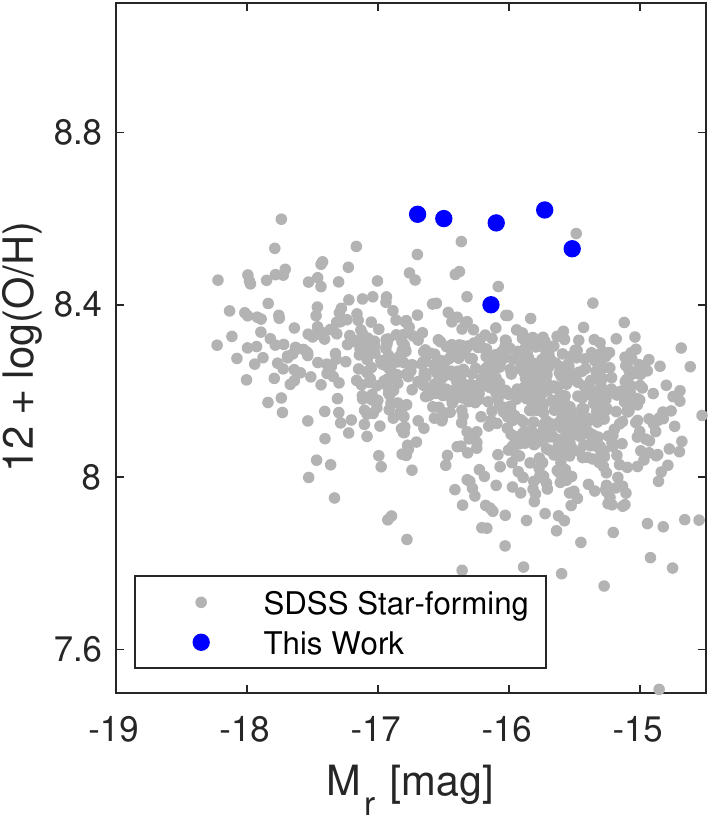}
\caption{Relation between emission line metallicity, 12 + log(O/H), and $r$-band absolute magnitude. Blue circles for our six sample galaxies. Grey dots are for the SDSS star-forming galaxies of the redshift range of 0.01\,$\sim$\,0.02.}
\label{masmet}
\end{figure*}

Analysis of the SDSS fiber spectroscopic data reveals that the nuclei have relatively high emission-line metallicity (12 + log(O/H) $>$ 8.4 dex), with a median 8.59. The solar value is 8.66 dex \citep{Pilyugin03}. In Figure \ref{masmet}, we show a relation between 12 + log(O/H) and the $r$-band absolute magnitude. The comparison sample galaxies are the SDSS star-forming galaxies of the redshift range of 0.01\,$\sim$\,0.02 \citep{Duc14}, and our galaxies significantly deviate. The derived metallicity of the underlying old stellar population seems to be consistent with that of typical dEs \citep{Paudel11} with a median value of log(Z/Z$_{\sun}$) being $-$0.8. This implies that the gas is about five times more metal-rich than the underlying old stellar population.

\section{Discussion}

\subsection{Star Formation}

The SFR derived from the SDSS fiber spectra is, on average, $\sim$\,0.03 M$_{\sun}$/yr. As we pointed out above, the 3\arcsec{} fiber mainly samples the central region, particularly the nucleus. We also measure the global SFRs of the galaxies from the FUV images observed with the Galaxy Evolution Explorer \citep{Martin05}. Interestingly, the FUV-based global SFRs are consistent with the SDSS fiber SFRs. This indicates that the star-formation in the galaxies is indeed concentrated within the nuclei region and the contribution from the outer part is minimal.

Examining the literature for radio observations of these dEs, we find that dE5 (UGCA 298) was detected by the ALFA-ALFA survey. It has a gas mass of 1.7$\times$10$^{7}$ M$_{\sun}$, approximately 10 times lower than its stellar mass. Assuming an HI consumption fraction of 10\%, the star-formation activity will last only a few Myr with the current SFR (0.025 M$_{\sun}$ yr$^{-1}$). It is thus very likely that dE5's star formation is temporal and at its final phase before turning into a true red and dead dE.

The underlying halo stellar population is clearly old, and another critical question to address is how the star-forming gas is acquired. All of the dEs we study here are located significantly away from the cluster/group centers (see Table \ref{sampletab}, column 11). Given the radial velocities are very similar to their host cluster/group central galaxies, these dEs might be the backsplash ones \citep{Jaffe15}. They have radially gone through the cluster/group core quickly and remained as backsplash galaxies at the outskirt of the cluster/group, where they acquire the gas leading to star formation \citep{Almeida14}. But the observed high metallicity does not favor this hypothesis. If they could have acquired fresh gas from the outskirts of a cluster/group, the observed emission-line metallicity would be much lower.

It has also been argued that star-formation activity can be rejuvenated from recycled gas that flows into the galactic center \citep{Boselli08}. That may explain the observed relatively high metallicity of these young star-forming blobs. A more quantitative conclusion can be drawn from detailed abundance analyses for old stellar populations that are not contaminated by the emission lines from gas. The absorption line indices are currently hard to measure precisely from the SDSS spectra because they are dominated by emission lines.

\subsection{Formation of Nuclei}

We have identified six relatively old dEs that host a compact star-forming blob near (or at) the center. Could these star-forming blobs be future nuclei as typically seen in dEs? Detailed studies of the stellar population properties of dE nuclei have revealed that the majority of them are younger than the halos of their host galaxies, but the origin of the young age remains elusive \citep{Paudel11}. Two main scenarios have been proposed: (1) in situ formation of nuclei in an early epoch of galaxy evolution \citep{Bergh86,Milosavljevic04}, and (2) reactivated star formation by the fueling of star-forming gas later on \citep{Bekki07}.  
 
It is also possible that the entire nuclear structure is formed in a later phase of galaxy evolution with a burst of star formation via the sinking of star-forming gas. That creates younger nuclei at the centers of galaxies. In this vein, \citet{Guillard16} proposed a `wet migration' scenario, where a massive star cluster could form a little away from the center of a galaxy and migrate to the center via a combination of interactions with other substructures and dynamical friction. Indeed, we have identified a few exemplary cases of such off-centered compact star-forming regions that might act as seeds of nuclei as proposed in the wet migration scenario.

\acknowledgments
We thank Florent Renaud and Rory Smith for fruitful discussions and comments on the draft version of this paper.
S.P. acknowledges support from the New Researcher Program (Shinjin grant No.\,2019R1C1C1009600) through the National Research Foundation of Korea. S.-J.Y. acknowledges support by the Mid-career Researcher Program (No.\,2019R1A2C3006242) and the SRC Program (the Center for Galaxy Evolution Research; No.\,2017R1A5A1070354) through the National Research Foundation of Korea. This study is based on the archival images and spectra from the Sloan Digital Sky Survey (the full acknowledgment can be found at http://www.sdss.org/collaboration/credits.html).


\begin{thebibliography}{}
\expandafter\ifx\csname natexlab\endcsname\relax\def\natexlab#1{#1}\fi
\providecommand{\url}[1]{\href{#1}{#1}}
\providecommand{\dodoi}[1]{doi:~\href{http://doi.org/#1}{\nolinkurl{#1}}}
\providecommand{\doeprint}[1]{\href{http://ascl.net/#1}{\nolinkurl{http://ascl.net/#1}}}
\providecommand{\doarXiv}[1]{\href{https://arxiv.org/abs/#1}{\nolinkurl{https://arxiv.org/abs/#1}}}

\bibitem[{{Abazajian} {et~al.}(2009){Abazajian}, {Adelman-McCarthy},
  {Ag{\"u}eros}, {Allam}, {Allende Prieto}, {An}, {Anderson}, {Anderson},
  {Annis}, {Bahcall}, \& et~al.}]{Abazajian09}
{Abazajian}, K.~N., {Adelman-McCarthy}, J.~K., {Ag{\"u}eros}, M.~A., {et~al.}
  2009, \apjs, 182, 543, \dodoi{10.1088/0067-0049/182/2/543}

\bibitem[{{Bekki}(2007)}]{Bekki07}
{Bekki}, K. 2007, \pasa, 24, 77, \dodoi{10.1071/AS07008}

\bibitem[{{Bekki} \& {Freeman}(2003)}]{Bekki03}
{Bekki}, K., \& {Freeman}, K.~C. 2003, \mnras, 346, L11,
  \dodoi{10.1046/j.1365-2966.2003.07275.x}

\bibitem[{{Bell} {et~al.}(2003){Bell}, {McIntosh}, {Katz}, \&
  {Weinberg}}]{Bell03}
{Bell}, E.~F., {McIntosh}, D.~H., {Katz}, N., \& {Weinberg}, M.~D. 2003, \apjs,
  149, 289, \dodoi{10.1086/378847}

\bibitem[{{B{\"o}ker}(2010)}]{Boker10}
{B{\"o}ker}, T. 2010, in IAU Symposium, Vol. 266, Star Clusters: Basic Galactic
  Building Blocks Throughout Time and Space, ed. R.~{de Grijs} \& J.~R.~D.
  {L{\'e}pine}, 58--63

\bibitem[{{Boselli} {et~al.}(2008){Boselli}, {Boissier}, {Cortese}, \&
  {Gavazzi}}]{Boselli08}
{Boselli}, A., {Boissier}, S., {Cortese}, L., \& {Gavazzi}, G. 2008, \apj, 674,
  742, \dodoi{10.1086/525513}

\bibitem[{{Cen}(2001)}]{Cen01}
{Cen}, R. 2001, \apj, 560, 592, \dodoi{10.1086/323071}

\bibitem[{{Chilingarian}(2009)}]{Chilingarian09}
{Chilingarian}, I.~V. 2009, \mnras, 394, 1229,
  \dodoi{10.1111/j.1365-2966.2009.14450.x}

\bibitem[{{C{\^o}t{\'e}} {et~al.}(2006){C{\^o}t{\'e}}, {Piatek}, {Ferrarese},
  {Jord{\'a}n}, {Merritt}, {Peng}, {Ha{\c s}egan}, {Blakeslee}, {Mei}, {West},
  {Milosavljevi{\'c}}, \& {Tonry}}]{Cote06}
{C{\^o}t{\'e}}, P., {Piatek}, S., {Ferrarese}, L., {et~al.} 2006, \apjs, 165,
  57, \dodoi{10.1086/504042}

\bibitem[{{den Brok} {et~al.}(2014){den Brok}, {Peletier}, {Seth}, {Balcells},
  {Dominguez}, {Graham}, {Carter}, {Erwin}, {Ferguson}, {Goudfrooij},
  {Guzm{\'a}n}, {Hoyos}, {Jogee}, {Lucey}, {Phillipps}, {Puzia}, {Valentijn},
  {Kleijn}, \& {Weinzirl}}]{Brok14}
{den Brok}, M., {Peletier}, R.~F., {Seth}, A., {et~al.} 2014, \mnras, 445,
  2385, \dodoi{10.1093/mnras/stu1906}

\bibitem[{{Duc} {et~al.}(2014){Duc}, {Paudel}, {McDermid}, {Cuillandre},
  {Serra}, {Bournaud}, {Cappellari}, \& {Emsellem}}]{Duc14}
{Duc}, P.-A., {Paudel}, S., {McDermid}, R.~M., {et~al.} 2014, \mnras, 440,
  1458, \dodoi{10.1093/mnras/stu330}

\bibitem[{{Ferrarese} {et~al.}(2006){Ferrarese}, {C{\^o}t{\'e}}, {Dalla
  Bont{\`a}}, {Peng}, {Merritt}, {Jord{\'a}n}, {Blakeslee}, {Ha{\c s}egan},
  {Mei}, {Piatek}, {Tonry}, \& {West}}]{Ferrarese06}
{Ferrarese}, L., {C{\^o}t{\'e}}, P., {Dalla Bont{\`a}}, E., {et~al.} 2006,
  \apjl, 644, L21, \dodoi{10.1086/505388}

\bibitem[{{Georgiev} \& {B{\"o}ker}(2014)}]{Georgiev14}
{Georgiev}, I.~Y., \& {B{\"o}ker}, T. 2014, \mnras, 441, 3570,
  \dodoi{10.1093/mnras/stu797}

\bibitem[{{Gu{\'e}rou} {et~al.}(2015){Gu{\'e}rou}, {Emsellem}, {McDermid},
  {C{\^o}t{\'e}}, {Ferrarese}, {Blakeslee}, {Durrell}, {MacArthur}, {Peng},
  {Cuilland re}, \& {Gwyn}}]{Guerou15}
{Gu{\'e}rou}, A., {Emsellem}, E., {McDermid}, R.~M., {et~al.} 2015, \apj, 804,
  70, \dodoi{10.1088/0004-637X/804/1/70}

\bibitem[{{Guillard} {et~al.}(2016){Guillard}, {Emsellem}, \&
  {Renaud}}]{Guillard16}
{Guillard}, N., {Emsellem}, E., \& {Renaud}, F. 2016, \mnras, 461, 3620,
  \dodoi{10.1093/mnras/stw1570}

\bibitem[{{Jaff{\'e}} {et~al.}(2015){Jaff{\'e}}, {Smith}, {Candlish},
  {Poggianti}, {Sheen}, \& {Verheijen}}]{Jaffe15}
{Jaff{\'e}}, Y.~L., {Smith}, R., {Candlish}, G.~N., {et~al.} 2015, \mnras, 448,
  1715, \dodoi{10.1093/mnras/stv100}

\bibitem[{{Jiang} {et~al.}(2011){Jiang}, {Greene}, {Ho}, {Xiao}, \&
  {Barth}}]{Jiang11}
{Jiang}, Y.-F., {Greene}, J.~E., {Ho}, L.~C., {Xiao}, T., \& {Barth}, A.~J.
  2011, \apj, 742, 68, \dodoi{10.1088/0004-637X/742/2/68}

\bibitem[{{Kacharov} {et~al.}(2018){Kacharov}, {Neumayer}, {Seth},
  {Cappellari}, {McDermid}, {Walcher}, \& {B{\"o}ker}}]{Kacharov18}
{Kacharov}, N., {Neumayer}, N., {Seth}, A.~C., {et~al.} 2018, \mnras, 480,
  1973, \dodoi{10.1093/mnras/sty1985}

\bibitem[{{Kennicutt}(1998)}]{Kennicutt98}
{Kennicutt}, Jr., R.~C. 1998, \araa, 36, 189,
  \dodoi{10.1146/annurev.astro.36.1.189}

\bibitem[{{Koleva} {et~al.}(2009){Koleva}, {Prugniel}, {Bouchard}, \&
  {Wu}}]{Koleva09}
{Koleva}, M., {Prugniel}, P., {Bouchard}, A., \& {Wu}, Y. 2009, \aap, 501,
  1269, \dodoi{10.1051/0004-6361/200811467}

\bibitem[{{Leitherer} {et~al.}(1999){Leitherer}, {Schaerer}, {Goldader},
  {Delgado}, {Robert}, {Kune}, {de Mello}, {Devost}, \&
  {Heckman}}]{Leitherer99}
{Leitherer}, C., {Schaerer}, D., {Goldader}, J.~D., {et~al.} 1999, \apjs, 123,
  3, \dodoi{10.1086/313233}

\bibitem[{{Lisker} {et~al.}(2006){Lisker}, {Glatt}, {Westera}, \&
  {Grebel}}]{Lisker06}
{Lisker}, T., {Glatt}, K., {Westera}, P., \& {Grebel}, E.~K. 2006, \aj, 132,
  2432, \dodoi{10.1086/508414}

\bibitem[{{Lisker} {et~al.}(2008){Lisker}, {Grebel}, \& {Binggeli}}]{Lisker08}
{Lisker}, T., {Grebel}, E.~K., \& {Binggeli}, B. 2008, \aj, 135, 380,
  \dodoi{10.1088/0004-6256/135/1/380}

\bibitem[{{Lisker} {et~al.}(2007){Lisker}, {Grebel}, {Binggeli}, \&
  {Glatt}}]{Lisker07}
{Lisker}, T., {Grebel}, E.~K., {Binggeli}, B., \& {Glatt}, K. 2007, \apj, 660,
  1186, \dodoi{10.1086/513090}

\bibitem[{{Lotz} {et~al.}(2004){Lotz}, {Miller}, \& {Ferguson}}]{Lotz04}
{Lotz}, J.~M., {Miller}, B.~W., \& {Ferguson}, H.~C. 2004, \apj, 613, 262,
  \dodoi{10.1086/422871}

\bibitem[{{Marino} {et~al.}(2013){Marino}, {Rosales-Ortega}, {S{\'a}nchez},
  {Gil de Paz}, {V{\'{\i}}lchez}, {Miralles-Caballero}, {Kehrig},
  {P{\'e}rez-Montero}, {Stanishev}, {Iglesias-P{\'a}ramo}, {D{\'{\i}}az},
  {Castillo-Morales}, {Kennicutt}, {L{\'o}pez-S{\'a}nchez}, {Galbany},
  {Garc{\'{\i}}a-Benito}, {Mast}, {Mendez-Abreu}, {Monreal-Ibero}, {Husemann},
  {Walcher}, {Garc{\'{\i}}a-Lorenzo}, {Masegosa}, {Del Olmo Orozco},
  {Mour{\~a}o}, {Ziegler}, {Moll{\'a}}, {Papaderos},
  {S{\'a}nchez-Bl{\'a}zquez}, {Gonz{\'a}lez Delgado}, {Falc{\'o}n-Barroso},
  {Roth}, {van de Ven}, \& {Califa Team}}]{Marino13}
{Marino}, R.~A., {Rosales-Ortega}, F.~F., {S{\'a}nchez}, S.~F., {et~al.} 2013,
  \aap, 559, A114, \dodoi{10.1051/0004-6361/201321956}

\bibitem[{{Martin} {et~al.}(2005){Martin}, {Fanson}, {Schiminovich},
  {Morrissey}, {Friedman}, {Barlow}, {Conrow}, {Grange}, {Jelinsky},
  {Milliard}, {Siegmund}, {Bianchi}, {Byun}, {Donas}, {Forster}, {Heckman},
  {Lee}, {Madore}, {Malina}, {Neff}, {Rich}, {Small}, {Surber}, {Szalay},
  {Welsh}, \& {Wyder}}]{Martin05}
{Martin}, D.~C., {Fanson}, J., {Schiminovich}, D., {et~al.} 2005, \apjl, 619,
  L1, \dodoi{10.1086/426387}

\bibitem[{{McLaughlin} {et~al.}(2006){McLaughlin}, {King}, \&
  {Nayakshin}}]{McLaughlin06}
{McLaughlin}, D.~E., {King}, A.~R., \& {Nayakshin}, S. 2006, \apjl, 650, L37,
  \dodoi{10.1086/508627}

\bibitem[{{Mei} {et~al.}(2007){Mei}, {Blakeslee}, {C{\^o}t{\'e}}, {Tonry},
  {West}, {Ferrarese}, {Jord{\'a}n}, {Peng}, {Anthony}, \& {Merritt}}]{Mei07}
{Mei}, S., {Blakeslee}, J.~P., {C{\^o}t{\'e}}, P., {et~al.} 2007, \apj, 655,
  144, \dodoi{10.1086/509598}

\bibitem[{{Mieske} {et~al.}(2013){Mieske}, {Frank}, {Baumgardt},
  {L{\"u}tzgendorf}, {Neumayer}, \& {Hilker}}]{Mieske13}
{Mieske}, S., {Frank}, M.~J., {Baumgardt}, H., {et~al.} 2013, \aap, 558, A14,
  \dodoi{10.1051/0004-6361/201322167}

\bibitem[{{Milosavljevi{\'c}}(2004)}]{Milosavljevic04}
{Milosavljevi{\'c}}, M. 2004, \apjl, 605, L13, \dodoi{10.1086/420696}

\bibitem[{{Oh} \& {Lin}(2000)}]{Oh00}
{Oh}, K.~S., \& {Lin}, D.~N.~C. 2000, \apj, 543, 620, \dodoi{10.1086/317118}

\bibitem[{{Paudel} {et~al.}(2014){Paudel}, {Lisker}, {Hansson}, \&
  {Huxor}}]{Paudel14}
{Paudel}, S., {Lisker}, T., {Hansson}, K.~S.~A., \& {Huxor}, A.~P. 2014,
  \mnras, 443, 446, \dodoi{10.1093/mnras/stu1171}

\bibitem[{{Paudel} {et~al.}(2010){Paudel}, {Lisker}, \& {Janz}}]{Paudel10}
{Paudel}, S., {Lisker}, T., \& {Janz}, J. 2010, \apjl, 724, L64,
  \dodoi{10.1088/2041-8205/724/1/L64}

\bibitem[{{Paudel} {et~al.}(2011){Paudel}, {Lisker}, \&
  {Kuntschner}}]{Paudel11}
{Paudel}, S., {Lisker}, T., \& {Kuntschner}, H. 2011, \mnras, 413, 1764,
  \dodoi{10.1111/j.1365-2966.2011.18256.x}

\bibitem[{{Pilyugin} {et~al.}(2003){Pilyugin}, {Ferrini}, \&
  {Shkvarun}}]{Pilyugin03}
{Pilyugin}, L.~S., {Ferrini}, F., \& {Shkvarun}, R.~V. 2003, \aap, 401, 557,
  \dodoi{10.1051/0004-6361:20030139}

\bibitem[{{Reines} {et~al.}(2013){Reines}, {Greene}, \& {Geha}}]{Reines13}
{Reines}, A.~E., {Greene}, J.~E., \& {Geha}, M. 2013, \apj, 775, 116,
  \dodoi{10.1088/0004-637X/775/2/116}

\bibitem[{{Rossa} {et~al.}(2006){Rossa}, {van der Marel}, {B{\"o}ker},
  {Gerssen}, {Ho}, {Rix}, {Shields}, \& {Walcher}}]{Rossa06}
{Rossa}, J., {van der Marel}, R.~P., {B{\"o}ker}, T., {et~al.} 2006, \aj, 132,
  1074, \dodoi{10.1086/505968}

\bibitem[{{S{\'a}nchez Almeida} {et~al.}(2014){S{\'a}nchez Almeida},
  {Elmegreen}, {Mu{\~n}oz-Tu{\~n}{\'o}n}, \& {Elmegreen}}]{Almeida14}
{S{\'a}nchez Almeida}, J., {Elmegreen}, B.~G., {Mu{\~n}oz-Tu{\~n}{\'o}n}, C.,
  \& {Elmegreen}, D.~M. 2014, \aapr, 22, 71, \dodoi{10.1007/s00159-014-0071-1}

\bibitem[{{S{\'a}nchez-Janssen} {et~al.}(2019){S{\'a}nchez-Janssen},
  {C{\^o}t{\'e}}, {Ferrarese}, {Peng}, {Roediger}, {Blakeslee}, {Emsellem},
  {Puzia}, {Spengler}, {Taylor}, {{\'A}lamo-Mart{\'\i}nez}, {Boselli},
  {Cantiello}, {Cuillandre}, {Duc}, {Durrell}, {Gwyn}, {MacArthur},
  {Lan{\c{c}}on}, {Lim}, {Liu}, {Mei}, {Miller}, {Mu{\~n}oz}, {Mihos},
  {Paudel}, {Powalka}, \& {Toloba}}]{Janssen19}
{S{\'a}nchez-Janssen}, R., {C{\^o}t{\'e}}, P., {Ferrarese}, L., {et~al.} 2019,
  \apj, 878, 18, \dodoi{10.3847/1538-4357/aaf4fd}

\bibitem[{{Seth} {et~al.}(2006){Seth}, {Dalcanton}, {Hodge}, \&
  {Debattista}}]{Seth06}
{Seth}, A.~C., {Dalcanton}, J.~J., {Hodge}, P.~W., \& {Debattista}, V.~P. 2006,
  \aj, 132, 2539, \dodoi{10.1086/508994}

\bibitem[{{Seth} {et~al.}(2014){Seth}, {van den Bosch}, {Mieske}, {Baumgardt},
  {Brok}, {Strader}, {Neumayer}, {Chilingarian}, {Hilker}, {McDermid},
  {Spitler}, {Brodie}, {Frank}, \& {Walsh}}]{Seth14}
{Seth}, A.~C., {van den Bosch}, R., {Mieske}, S., {et~al.} 2014, \nat, 513,
  398, \dodoi{10.1038/nature13762}

\bibitem[{{Theureau} {et~al.}(2007){Theureau}, {Hanski}, {Coudreau}, {Hallet},
  \& {Martin}}]{Theureau07}
{Theureau}, G., {Hanski}, M.~O., {Coudreau}, N., {Hallet}, N., \& {Martin},
  J.-M. 2007, \aap, 465, 71, \dodoi{10.1051/0004-6361:20066187}

\bibitem[{{Tremaine} {et~al.}(1975){Tremaine}, {Ostriker}, \&
  {Spitzer}}]{Tremaine75}
{Tremaine}, S.~D., {Ostriker}, J.~P., \& {Spitzer}, Jr., L. 1975, \apj, 196,
  407, \dodoi{10.1086/153422}

\bibitem[{{Turner} {et~al.}(2012){Turner}, {C{\^o}t{\'e}}, {Ferrarese},
  {Jord{\'a}n}, {Blakeslee}, {Mei}, {Peng}, \& {West}}]{Turner12}
{Turner}, M.~L., {C{\^o}t{\'e}}, P., {Ferrarese}, L., {et~al.} 2012, \apjs,
  203, 5, \dodoi{10.1088/0067-0049/203/1/5}

\bibitem[{{Urich} {et~al.}(2017){Urich}, {Lisker}, {Janz}, {van de Ven},
  {Leaman}, {Boselli}, {Paudel}, {Sybilska}, {Peletier}, {den Brok}, {Hensler},
  {Toloba}, {Falc{\'o}n-Barroso}, \& {Niemi}}]{Urich}
{Urich}, L., {Lisker}, T., {Janz}, J., {et~al.} 2017, \aap, 606, A135,
  \dodoi{10.1051/0004-6361/201730897}

\bibitem[{{van den Bergh}(1986)}]{Bergh86}
{van den Bergh}, S. 1986, \aj, 91, 271, \dodoi{10.1086/114006}

\bibitem[{{Vazdekis} {et~al.}(2010){Vazdekis}, {S{\'a}nchez-Bl{\'a}zquez},
  {Falc{\'o}n-Barroso}, {Cenarro}, {Beasley}, {Cardiel}, {Gorgas}, \&
  {Peletier}}]{Vazdekis10}
{Vazdekis}, A., {S{\'a}nchez-Bl{\'a}zquez}, P., {Falc{\'o}n-Barroso}, J.,
  {et~al.} 2010, \mnras, 404, 1639, \dodoi{10.1111/j.1365-2966.2010.16407.x}

\bibitem[{{Voggel} {et~al.}(2016){Voggel}, {Hilker}, \& {Richtler}}]{Vogge16}
{Voggel}, K., {Hilker}, M., \& {Richtler}, T. 2016, \aap, 586, A102,
  \dodoi{10.1051/0004-6361/201527070}

\bibitem[{{Zhang} {et~al.}(2017){Zhang}, {Puzia}, \& {Weisz}}]{Zhang17}
{Zhang}, H.-X., {Puzia}, T.~H., \& {Weisz}, D.~R. 2017, \apjs, 233, 13,
  \dodoi{10.3847/1538-4365/aa937b}

\end{thebibliography}

\begin{table*}
\caption{Global properties of sample galaxies}
\scriptsize
\begin{tabular}{|ccccccccccccccc|}
\hline
Galaxy & R.A. & Dec. & M$_{B}$ & z  & $g-r$ & $R_{e}$ & n & FUV (SFR) & M$^{*}$  & R & R.V. & Env. \\
 & (h:m:s) & (d:m:s) & (mag) & & (mag) & (kpc) &  & (mag)\,\,\,\trev{(log[M$_{\sun}\,$yr$^{-1}$])} & log(M$_{\sun}$) & (Mpc) & (km\,s$^{-1}$) &\\
(1) & (2) & (3) & (4) & (5) & (6) & (7) & (8) & (9) & (10) & (11) & (12) & (13) \\
\hline
DE1 & 11:17:17.02 & +16:19:37.65 & $-$15.60 & 0.0036 & 0.89 &  12.5 & 1.1 & 19.7 ($-$2.11) & 8.31   &  5.4 &  204 &  Virgo  \\
DE2 & 11:30:26.21 & +17:19:56.69 & $-$15.02 & 0.0039 & 0.94 &  06.4 & 1.5 & 19.9 ($-$2.62) & 8.19   &  4.5 &  114 &  Virgo  \\
DE3 & 12:24:22.18 & +21:09:35.95 & $-$16.20 & 0.0030 & 0.81 &  12.7 & 1.3 & 19.0 ($-$1.93) & 8.27   &  2.5 &  384 &  Virgo  \\
DE4 & 12:40:50.32 & +04:31:32.99 & $-$15.64 & 0.0025 & 1.01 &  11.7 & 1.0 & 19.1 ($-$1.85) & 8.20   &  2.3 &  534 &  Virgo  \\
DE5 & 12:46:55.40 & +26:33:51.38 & $-$16.00 & 0.0027 & 0.94 &  09.1 & 2.2 & 18.8 ($-$1.70) & 8.35   &  4.2 &  474 &  Virgo  \\
DE6 & 11:53:49.00 & +60:52:09.42 & $-$15.23 & 0.0041 & 0.96 &  08.2 & 1.2 & 19.9 ($-$2.52) & 8.20   &  0.7 &  155 &  N3945  \\
\hline
\end{tabular}
\tablecomments{The magnitudes and redshifts are obtained from the SDSS catalog. The $B$-band magnitudes are converted from the $g$-band magnitudes using an equation, M$_B$ = M$_g$ + 0.313 $\times$ ($g-r$) + 0.227, obtained from the SDSS web-page\footnote{http://www.sdss3.org/dr8/algorithms/sdssUBVRITransform.php\#Lupton2005}. The structural parameters ($R_{e}$ and n) are derived from surface photometry in the SDSS $i$-band image excluding central nuclei. The $g-r$ colors are measured after masking the central 5\arcsec{} regions. We perform aperture photometry on publicly available GALEX \citep{Martin05} all-sky survey images to measure FUV fluxes listed in column 9 where the bracketed values are log(SFR) based on FUV fluxes. In column 10, we list stellar masses of the galaxies derived from $r$-band magnitudes. In the columns 11 and 12, we provide sky-projected distances (R) and relative line-of-sight velocities (R.V.) of the galaxies from the cluster center where they reside (column 13).}
\label{sampletab}
\end{table*}

\begin{table*}
\caption{Properties of young nuclei}
\footnotesize
\begin{tabular}{|c|ccccc|ccccc|cc|}
\hline
& \multicolumn{5}{c|}{Image analysis } & \multicolumn{5}{c|}{Emission line analysis} & \multicolumn{2}{c|}{Old stellar population}  \\
\hline
Galaxy & M$_{r}$  & $g-r$ &  log(M$_{*}$) & $\Delta$R & FWHM & 12+log(O/H) & SFR & $c$ & EW(H$_{\alpha}$) & Age  &   Age        &      log(Z/Z$_{\sun}$)  \\
& (mag) & (mag) & (M$_{\sun}$) &  (\arcsec)  & (\arcsec) & (dex) & \trev{(log[M$_{\sun}$\,yr$^{-1}$])} & & (\AA) & (Myr)&  (Gyr) & (dex)\\
(1)&  (2) & (3) & (4) & (5) & (5)  & (7) &  (8)  & (9)  &  (10)  & (11)  &  (12) &   (13) \\
\hline
DE1 &  $-$13.3 & $-$0.4 & 6.70 & 3.2 & 0.8  & 8.59 &  $-$1.99  & 3.4  &  319  & 5.04  &  $-$ &   $-$ \\
DE2 &  $-$12.5 & $-$0.3 & 6.45 & 0.8 & 1.0  & 8.53 &  $-$2.20  & 3.0  &  111  & 6.21  &  0.4$\pm$0.1 &   $-$0.6$\pm$0.02 \\
DE3 &  $-$12.7 & $-$0.5 & 6.40 & 2.2 & 1.1  & 8.61 &  $-$2.16  & 3.3  &  124  & 6.13  &  1.2$\pm$0.1 &   $-$1.3$\pm$0.05 \\
DE4 &  $-$13.1 & $-$0.3 & 6.69 & 0.7 & 1.2  & 8.40 &  $-$2.31  & 3.4  &  83  & 6.24  &  0.7$\pm$0.1 &   $-$0.5$\pm$0.02 \\
DE5 &  $-$13.5 & $-$0.5 & 6.72 & 0.3 & 0.9  & 8.60 &  $-$1.60  & 3.5  &  199  & 5.61  &  9.8$\pm$1.1 &   $-$1.0$\pm$0.02 \\
DE6 &  $-$13.1 & $-$0.5 & 6.56 & 1.3 & 1.3  & 8.62 &  $-$1.90  & 3.0  &  384  & 4.89  &  $-$ &   $-$ \\
\hline
\end{tabular}
\tablecomments{We list $r$-band absolute magnitudes, $g-r$ color, and stellar masses in columns 2, 3, and 4, respectively. The stellar mass is derived from $r$-band luminosity with a mass-to-light ratio calculated from $g-r$ based on \cite{Bell03}. In columns 5 and 6, we provide the galactocentric distances of the nuclei ($\Delta$R) and the observed seeing of the SDSS $g$-band (FWHM), respectively. Columns 7--11 are for the result of spectroscopic data analysis. The last two columns are the ages and metallicities of the underlying old stellar population of host galaxies derived from fitting of full spectra.}
\label{nuctab}
\end{table*}

\end{document}